# Refractive index of vanadium determined by resonant diffraction of soft x-rays

Martin Magnuson and Coryn F. Hague

*Université Pierre et Marie Curie (Paris VI), Laboratoire de Chimie Physique - Matière et Rayonnement (UMR 7614), 11 rue Pierre et Marie Curie, F-75231 Paris Cedex 05, France.*

**Abstract**

The dispersive part of the refractive index of vanadium is determined by measuring the angular displacement of the first order diffraction peak of a V/Fe superlattice. The measurements were made using elliptically polarized synchrotron radiation which was scanned through the V $L_{2,3}$ absorption edges for different incident scattering angles. The x-ray scattering technique provides access to direct determination of the dispersive part of the refractive index through an absorption resonance. The influence of absorption at the resonances is shown by comparing the absorption correction to the dispersion correction. The results demonstrate that 1-δ is larger than unity at the $L_{2,3}$ resonances of vanadium and the optical consequences are discussed.

## 1 Introduction

In the hard x-ray energy regime, deviations from Bragg's law of diffraction were already discovered in 1919 in the vicinity of absorption resonances of crystals[1]. These deviations indicated the existence of a refractive index for x-rays, whose real part was slightly less than unity since the x-rays were refracted in a direction slightly away from the surface normal[2]. However, as we know today this is not always strictly valid close to absorption resonances where the refractive index occasionally becomes larger than unity. Further investigations of the phenomenon of total external reflection at glancing incidence below the so-called critical angle were made as a result of the small but measurable refraction of the x-rays from the surface normal[3]. Precise measurements of the so-called anomalous dispersion effect of the refractive index in the vicinity of deep absorption resonances made it possible to estimate the refractive index in various crystals[4].

Detailed knowledge of the dispersive and absorptive parts of the refractive index, or alternately anomalous scattering factors, are indispensable to interpret diffraction and reflectivity data recorded in the context of multilayers and nano-engineered synthetic thin film materials close to absorption edges[5]. In particular, precise measurements of the refractive index are essential in the soft x-ray region where the greater absorption resonances cause strong variations of the refractive index. Although soft x-ray wavelengths are too long to produce diffraction peaks in single crystals, they are suitable for larger periodic structures such as multilayers with periodic spacings of a few monolayers. Layered materials of sufficient quality are commonly applied as reflecting, diffracting or dispersing elements and optics such as mirrors and gratings. Tables covering the x-ray region[6] for all elements present the absorptive part of the refractive index obtained directly from transmission measurements of thin foils from the photoabsorption coefficients. The dispersive part is obtained by application of the Kramers-Kronig transformation[7]. In the soft x-ray energy region it is often difficult to make sufficiently thin, free-standing films for transmission measurements. For this purpose, x-ray absorption spectroscopy from bulk samples using photoelectron yield techniques is often applied[8]. However, the relatively small depth probed by the absorption technique, limited to a few tens of Å is not suitable for studying thick multilayers and buried interfaces. It is also important to compare different experimental methods in addition to the standard Kramers-Kronig approach. Another element specific probe is resonant Bragg diffraction for which tunable synchrotron radiation is required. By scanning the photon energy across absorption resonances, Bragg scattering from a periodic structure such as a superlattice makes it possible to obtain the dispersive part of the atomic scattering factor[9], or, alternately, the refractive index without using the Kramers-Kronig transformation. In Ref. [10], this method was applied to a V/Fe superlattice and the dispersive part of the refractive index of Fe was extracted at the $2p-3d$ resonance.

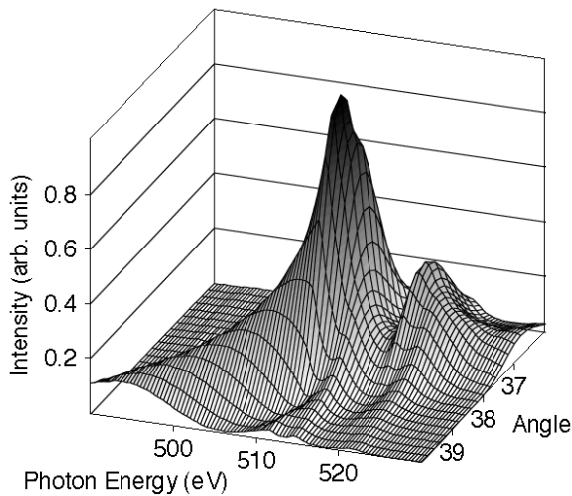

**Figure 1:** The first order modulated Bragg peak measured by scanning the photon energy through the V $L_{2,3}$ absorption resonances for various diffraction angles.

In order to gain further insight into the resonant behavior of refractive indices, we have pursued these ideas and performed x-ray Bragg diffraction measurements around the V $2p$ core-level thresholds of a V/Fe superlattice. The recent advances in the manufacturing of multilayers with a periodic spacing of a few Å are indispensable for producing suitable Bragg peaks in the soft x-ray energy regime. We take full advantage of the bulk sensitivity and element selectivity of the x-ray Bragg diffraction technique. It is shown how the displacement of the position of the Bragg peak maximum and the large variation of the intensity at the resonances makes it convenient to extract the dispersive part of the refractive index around the absorption edges. In the soft x-ray energy region where the absorption is much higher than in the hard x-ray region, the necessity of applying absorption correction at core-level resonances is investigated.

## 2  Experimental Details

The x-ray Bragg diffraction measurements were performed using the reflectometer at the soft x-ray metrology bending magnet beamline 6.3.2 at the Advanced Light Source (ALS)[11, 12]. Elliptically polarized radiation was used, by blanking-off part of the beam. The polarization rate was estimated to be ~ 60 %. The monochromator was set to a resolving power of about 2000 for a flux of ~ $10^{10}$ photons/second on the sample at the V $L_3$-edge.

The sample was epitaxially grown by dual-target magnetron sputtering deposition of metallic V and Fe layers on a polished MgO(001) fcc single crystal substrate at 300°C[13]. The alternating depositions of the V and Fe layers were repeated to form a total of 40 periods and capped with a Pd film to prevent oxidation. The structural quality of the sample was characterized by conventional $\tau$-$2\tau$ x-ray diffraction (XRD) measurements with Cu K$_\alpha$ radiation for low angles (1-14$^\circ$ in $2\tau$) and high angles (50-80$^\circ$ in $2\tau$) around the Fe/V (002) Bragg peak. The thickness parameters were obtained by a refinement procedure to reproduce the Bragg peaks of the XRD data using the simulation program SUPREX[14]. The periodicity $\Lambda = t_1 + t_2$ was determined to be $\sim$ 19.5 (19.78) Å and the individual thicknesses of Fe $\sim$ 7.5 Å (6 ML) and V $\sim$ 12 Å (7 ML). From the XRD data, the Pd capping layer was estimated to be $\sim$ 41 Å thick. The division parameter defined as the relative thickness ratio with respect to the periodicity: $\gamma = t_1/(t_1+t_2)$ was 0.6.

## 3 Results and Discussion

Figure 1 shows the resonant $L_{2,3}$ first order diffraction peak of vanadium measured both as a function of incident angle and photon energy, normalized to the incident photon flux. Strong variations in the intensity and width of the Bragg peak is observed when approaching the V $L_3$ edge at 514.0 eV while the $L_2$ resonance at 520.8 eV is weaker. Refraction changes the angle of propagation of the radiation entering the multilayer and therefore changes the angular positions of the Bragg peak maximum. The line shapes are asymmetric which indicates that absorption effects are important. At the $L_3$ resonance which has a peak maximum at $\sim$ 37.51$^\circ$, the width is significantly broader (2.03 eV) than the off-resonance value of $\sim$ 1.34 eV at 500 eV photon energy. The increased broadening at the resonance is a direct consequence of the increased absorption which reduces the number of participating superlattice planes scattering in phase.

Figure 2 shows the angular displacement of the position of the Bragg peak maximum as a function of photon energy. The characteristics of the Bragg peak in terms of position, width and intensity were extracted by standard interpolation and fitting procedures.

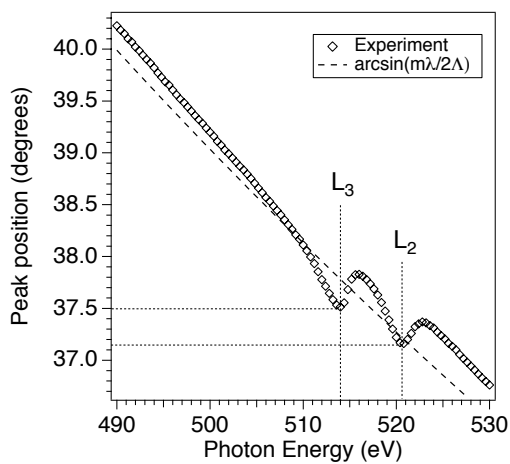

**Figure 2:** Angular displacement of the Bragg peak maximum as a function of photon energy at the 2$p$ edges of vanadium. The tilting dashed line is obtained from Bragg's law. The dotted lines indicate the angular positions of the intensity maxima at 37.51$^\circ$ at 514.0 eV and 37.16$^\circ$ at 520.8 eV at the $L_3$ and $L_2$ thresholds, respectively.

The dashed line represents the wavelength dependence of Bragg's law; $\tau_B = \arcsin(p\lambda/2\Lambda)$, where $p$ is the order of diffraction, $\lambda$ is the wavelength and $\Lambda$ is the periodicity of the multilayer. The measured angular displacement of the Bragg peak position does not just depend on the change in wavelength. The deviations from the straight line around the $L_{2,3}$ resonances are directly related to the energy dependence of the refractive index decrement $\delta$. Assuming that the absorption is negligible ($\beta=0$), application of Snell's law gives the

Bragg equation corrected for the refractive index [15],

$$p\lambda = 2\Lambda \sin\tau \sqrt{(1 - \frac{2\bar{\delta} - \bar{\delta}^2}{\sin^2\tau})} \qquad (1)$$

where $\bar{\delta} = \gamma\delta_1 + (1-\gamma)\delta_2$ denotes the averaged dispersive part of the refractive index decrement of the two layers and $\gamma$ is the division parameter defined by the thickness ratio of the bilayers with respect to the periodicity. If $\Lambda$ is known, measurements of the angular positions of the Bragg peaks as a function of energy may be used to determine $\bar{\delta}$. If the individual thicknesses $t_1$ and $t_2$ and thus the different compositions are known, the decrement of the dispersive parts of the refractive indices $\delta_1$ and $\delta_2$ of the component materials can also be determined. Since $\bar{\delta} \ll 1$ in the soft x-ray region, the quadratic $\bar{\delta}^2$-correction term in Eqn. can safely be neglected which implies that the expression can be simplified to

$$\delta_1 = \left[\sin\tau(\sin\tau - \frac{p\lambda}{2\Lambda}) - \delta_2\right]/\gamma + \delta_2 \qquad (2)$$

A more advanced correction term is needed in order to take into account absorption at the Bragg peak position. Correcting for both dispersion and absorption effects, the full refractive correction can written as[16, 17],

$$\delta_1 = \left[\sin\tau(\sin\tau - \frac{p\lambda}{2\Lambda}) - \delta_2\right]/D + \delta_2 \qquad (3)$$

where

$$D = \gamma - \frac{(\beta_1 - \beta_2)\sin^2(p\pi\gamma)P^2(\tau_B)}{p^2\pi^2[\gamma\beta_1 + (1-\gamma)\beta_2]} \qquad (4)$$

with the polarization factors $P^2(\tau_B) = C$ or $[1+C-\cos^2(2\tau)]/[1+C]$, depending on whether the incident photon beam is S or P polarized, respectively[18]. The factor D (Eqn. 4) describes the absorption in terms of a deviation from the nominal division parameter $\gamma$. The constant C defines the small but non-negligible absorption with S-polarization. C=0 means no absorption correction in S-polarization.

Figure 3 shows the dispersive part of the refractive index (1-$\delta$) obtained from the two corrections; from Eqn. 2 and from Eqns. 3 and 4, respectively. Tabulated values by Henke *et al.*[6] are shown for comparison. For the absorption correction, an $L_{2,3}$ absorption spectrum[20] of V was normalized to the tabulated values over the energy region 490-530 eV. Since there are no absorption resonances in this energy region for Fe, nonresonant tabulated values were used for both $\delta$ and $\beta$ in this case. The polarization ratio R=(S-P)/(S+P)=0.6 of the elliptically polarized x-rays was taken into account by the weight factors 0.8 and 0.2 for the S and P polarizations, respectively. The constant C was chosen to be 0.0625 based on calculations using the Fresnel equations which yield a polarization ratio between the S and P polarizations of $\sim$ 0.87 in reflectivity at the Bragg peak position[19].

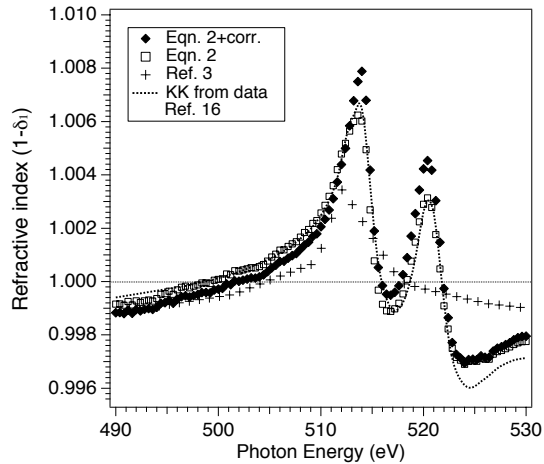

**Figure 3:** Photon energy dependence of the refractive index (1-δ) at the $L_{2,3}$ edges of vanadium obtained from measuring the angular displacement of the Bragg peak maximum. Literature values for bulk V (crosses) are shown for comparison[6].

In general, below the $L_{2,3}$ absorption edges, the refractive index is less than unity and is monotonically and slowly increasing in the effect known as normal dispersion from the terminology of visible light[21]. At about 501 eV, the refractive index becomes larger than unity and exponentially increases up to a peak value of 1.0065 at the $L_3$ edge at 513.6 eV and 1.0033 at 520.4 eV at the $L_2$ edge. After each absorption edge, the refractive index is strongly resonantly reduced back to a value less than unity in the effect known as anomalous dispersion.

The agreement with the tabulated (1-δ)-values in the literature[6] is reasonably good considering the fact that the tabulated values are for bulk V and the $2p$ spin-orbit splitting is not taken into account. The comparison with bulk values of the optical constants are justified since the V and Fe layers are relatively thick. The difference is largest for the energy range ~523-530 eV where the strong antiresonance effect in the experimental data is not taken into account in the tabulated values.

The absorption correction obtained from Eqns. 3 and 4 is found to have a very small effect on the refractive index 1-δ of vanadium. This is due to the fact that the incident elliptically polarized radiation is mainly S-polarized. Above and below the $L_{2,3}$ peak maxima, the (1-δ)-values obtained with the absorption correction are somewhat lower than those obtained by Eqn. 2 while at the peak positions, they are higher. The magnitude of the absorption correction thus strongly depends on the degree of polarization and is much larger for P-polarized than for S-polarized x-rays. In particular, this is emphasized in the present case since the Bragg angles (36.8-40.2$^O$) are close to the Brewster angle at ~ 44.9$^O$.

For x-ray optics such as mirrors, it is often desirable to maximize the critical angles and therefore absorption thresholds are generally being avoided. However, tuning grazing incident photons to energies close to an absorption threshold where 1-δ is larger than unity induce total internal reflection and standing waves along the interface of a multilayer. Like its visible counterpart, total internal reflection of x-rays can be anticipated to be utilized for turning x-ray beams within a limited bandwidth of photon energy.

# 4 Conclusions

Measurements of the first order modulation Bragg peak from a V/Fe multilayer is used to determine the dispersive part of the refractive index 1-δ around the 2*p* absorption thresholds of vanadium. Absorption correction was found to be small for elliptically polarized x-rays dominated by S-polarization. The results demonstrate that the dispersive part of the refractive index of V is greater than unity over the energy range around the $L_{2,3}$ absorption thresholds which has dramatic effects on the reflection properties.

# 5 Acknowledgments

We would like to thank P. Blomqvist for making the sample, M. Sacchi, E. Gullikson and J. Underwood for valuable assistance during measurements and the rest of the staff at the Advanced Light Source for making these measurements possible. This work was supported by the Swedish Foundation for International Cooperation in Research and Higher Education (STINT).

coefficient as 1 and $\cos^2(2\tau)$ for the S and P-polarizations, respectively. For absorption correction, the polarization factors have been modified to obtain larger absorption for P-polarization than S-polarization.